\documentclass[]{spie}  

\usepackage{subfig}
\usepackage{amsmath,amsfonts,amssymb}
\usepackage{graphicx}
\usepackage[colorlinks=true, allcolors=blue]{hyperref}
\usepackage{mwe}

\usepackage[inline]{enumitem}

\title{Status of predictive wavefront control on Keck II adaptive optics bench: on-sky coronagraphic results}

\author[a]{Maaike A.M. van Kooten}
\author[a]{Rebecca Jensen-Clem}
\author[b,c]{Sylvain Cetre}
\author[b]{Sam Ragland}
\author[d]{Charlotte Z. Bond}
\author[a]{J. Fowler}
\author[b]{Peter Wizinowich}
\affil[a]{University of California Santa Cruz, 156 High Street
Santa Cruz, CA, USA}
\affil[b]{W. M. Keck Observatory, 65-1120 Mamalahoa Hwy, Waimea, HI, USA}
\affil[c]{University of California 
Observatories,
1156 High Street
Santa Cruz, CA, USA }
\affil[d]{The UK Astronomy Technology Centre, Royal Observatory Edinburgh, Edinburgh, UK}

\authorinfo{Further author information: (Send correspondence to Maaike A.M van Kooten)\\ E-mail: mvankoot@ucsc.edu  }

\begin{document} 
\maketitle

\begin{abstract}
The behavior of an adaptive optics (AO) system for ground-based high contrast imaging (HCI) dictates the achievable contrast of the instrument. In conditions where the coherence time of the atmosphere is short compared to the speed of the AO system, the servo-lag error becomes the dominate error term of the AO system. While the AO system measures the wavefront error and subsequently applies a correction (taking a total of 1 to 2 milli-seconds), the atmospheric turbulence above the telescope has changed. In addition to reducing the Strehl ratio, the servo-lag error causes a build-up of speckles along the direction of the dominant wind vector in the coronagraphic image, severely limiting the contrast at small angular separations. One strategy to mitigate this problem is to predict the evolution of the turbulence over the delay. Our predictive wavefront control algorithm minimizes the delay in a mean square sense and has been implemented on the Keck II AO bench. In this paper we report on the latest results of our algorithm and discuss updates to the algorithm itself. We explore how to tune various filter parameters on the basis of both daytime laboratory tests and on-sky tests. We show a reduction in residual-mean-square wavefront error for the predictor compare to the leaky integrator implemented on Keck. Finally, we present contrast improvements for both day time and on-sky tests.  Using the L-band vortex coronagraph for Keck’s NIRC2 instrument, we find a contrast gain of 2.03 at separation of 3 $\lambda/D$ and up to 3 for larger separations (4-6$\lambda/D$). 

\end{abstract}


\section{INTRODUCTION}
\label{sec:intro}  

Single conjugate adaptive optics (AO) systems provide a good correction over a modest field-of-view, thereby enabling a vast array of science. Within a high contrast imaging (HCI) system, this simplest form of AO becomes a powerful tool when observing with bright, on-axis guide stars. Extreme AO correction is achieved by operating at high speeds and maximizing the actuator density of the deformable mirror (DM). The resulting systems provide Strehl ratios (SR) of greater than 60\% in near-infrared (NIR) wavelengths~\cite{Bond_2020}. 

With advancements in AO, HCI systems have reached post-processed contrasts of $10^{-6}$ at spatial separations of 200 milli-arc-seconds~\cite{zurlo_2016}. Closer to the host star, a typical AO system is dominated by either non-common path aberrations or the servo-lag error. The servo-lag error can result in a wind-driven halo for instruments such as Very Large Telescope's (VLT) SPHERE HCI instrument, contaminating the region near the inner working angle of the coronagraph~\cite{Cantalloube_2018}. The servo-lag error is due to the atmospheric turbulence above the telescope evolving in the time between when the wavefront is measured and the the new DM commands are applied. The delay time is calculated to include the wavefront sensor camera read-out time, DM response time, computation time, and 0.5 times the wavefront sensor (WFS) integration time; it ranges from 1-4 wavefront sensor frames, depending on the system. One solution to minimize the servo-lag error is to predict the evolution of the atmospheric turbulence over the known delay using predictive control. 

Over the last five years, efforts to implement predictive control on-sky for HCI have resulted in algorithms such as empirical orthogonal functions (EOF)~\cite{Guyon_2017} which is being tested at SCExAO~\cite{Nem_2015,cacao} and W.M.Keck~\cite{Jensen_2019}. Van Kooten et al. 2017, 2019~\cite{Maaike_2017,vanKooten_2019} proposed a recursive solution to the minimum mean square cost function and within their framework test prediction for non-stationary turbulence as well VLT/SPHERE telemetry~\cite{vanKooten_2020}. Most recently, data-driven subspace predictive control~\cite{Haffert_2021} has been proposed to operate in closed-loop, with plans to test on-sky with the MagAOX system~\cite{MagAOX}. 

In this paper, we focus on W.M. Keck observatory's HCI system, the Keck II's near-infrared camera (NIRC2). In Xuan et al. 2018~\cite{2018AJ....156..156X}, the authors found that the final post-processed contrast for NIRC2 is correlated with the coherence time of the atmosphere divided by the WFS integration time; both variables influence the impact of the servo-lag error. These results show that the performance of the system would be improved with the implementation of predictive control. Predictive control on NIRC2 was implemented and a factor of 2.2 reduction in the residual-mean-square (RMS) error was shown on the Keck II AO bench~\cite{Jensen_2019} during daytime tests with simulated turbulence projected onto the DM. Since then, updates to the AO system have been made along with the predictive controller. In this paper we present the latest results from daytime tests and on-sky engineering time at Keck including a new recursive implementation tested during the day and contrast measurements from both day and night time tests.  

We outline the paper as follows: an overview of Keck II AO and the NIRC2 instrument is given in Sec.~\ref{sec:Keck_AO_overview}. In Sec.~\ref{sec:controller}, we give details of the controller implemented on Keck including the predictive controller (Sec.~\ref{sec:pred_control}). We present results from daytime and nighttime tests in Secs.~\ref{sec:daytime} and~\ref{sec:nighttime}, respectively. We outline our future plans for the system in Sec.~\ref{sec:future}. Finally, we end with our summary and conclusions in Sec.~\ref{sec:conclusions}. 

\section{Keck II Adaptive Optics system and NIRC2}
\label{sec:Keck_AO_overview}  
NIRC2 (PI: Keith Matthews) contains two vortex coronagraphs to cover the full wavelength range of the instrument (K-,and L'-/M-band vortex, respectively) to enable HCI~\cite{L_band_vortex}. NIRC2 is fed by the Keck II AO bench that operates with either a visible Shack-Hartmann wavefront sensor (SHWFS) or a near-infrared pyramid wavefront sensor (PyWFS)~\cite{Bond_2018,Bond_2020}. The PyWFS enables AO correction for much fainter, redder guide stars enabling unique science capabilities. For the remainder of this paper we will be referring to the PyWFS configured AO system. 

The PyWFS controls the 21x21 actuator Xinetics deformable mirror (DM) having 40 pixels across the pupil, and operating with a modulation radius of 5 $\frac{\lambda}{D}$. The real-time computer (RTC) for the PyWFS makes use of the CACAO framework~\cite{cacao,Cetre_2018} allowing the system to operate at 1kHz. Simulations and on-sky commissioning have verified that the PyWFS is able to achieve upwards of 65~\%  Strehl ratio (SR) for an H-band magnitude of 5 and can provide AO correction up to H-band magnitude 12~\cite{Bond_2020}. The AO delay is determined to be 1.7 milli-seconds~\cite{Cetre_2018} when running the PyWFS at its full frame-rate. 
\section{Adaptive Optics Control}
\label{sec:controller}
\subsection{Integral control law}
Operating at 1kHz with 21 actuator across the aperture in H-band allows for the PyWFS to achieve good performance with the standard control law: a leaky integrator. In Eq.~\ref{eq:leaky_int}, the leaky integrator calculates the DM commands, $V^{int}(t)$, at time step $t$. The leak, $k$, is nominally set to 0.99 for standard operation and for our tests. The gain ($g$) has a typical value of 0.4 which is adjusted depending on seeing conditions. $S(t)$ and $S_{ref}$ are the slopes measured by the pyramid at time $t$ and the reference slopes encoding the non-command path aberrations for NIRC2, respectively. Finally, the command matrix, $CM$, not only converts the slopes to DM actuator voltage but also encodes the number of modes being used and any high- and low- order (LO and HO, respectively) modal gains. The modal gains are applied in Fourier space with the cut-off-frequency (CoF) indicating where the trade off between LO and HO modes occurs in units of $\lambda/D$. Typically, it is taken to be the modulation radius of the PyWFS. Unless specifically stated the high- and low-order modes were set to 0.4 and 1 respectively and the CoF is equal to 5. Finally, the total number of controlled Zernike modes can be varied as well, with 350 modes being the maximum correctable number of modes.  For more details on the controller and the slopes computation refer to Bond et. al. 2020~\cite{Bond_2020}.

\begin{equation}
    V^{int}(t)=k V^{int}(t-1)-g(CM \times [S(t)-S_{ref}])
    \label{eq:leaky_int}
\end{equation}

\subsection{Predictive Control}
\label{sec:pred_control}
Using the separation principle, our predictive controller is implemented in two steps: 1. predict the input disturbance (i.e, atmospheric turbulence above the telescope) 2. control the system plant (i.e., DM). Within this flexible framework, we can implement different predictive control methods. We are also able to pick various control laws including open-loop control or the already existing leaky integrator implemented for the PyWFS. 

In this work, we predict the atmospheric turbulence over the lag of the system by finding a filter that uses a subset of previous measurements subject to some cost function. Since we aim to predict the turbulence itself, we must first reconstruct the pseudo-open loop (POL) turbulence. 

\subsubsection{Pseudo-open loop reconstruction}
Making use of the shared memory structure within the CACAO RTC framework, the POL DM commands are calculated using the current PyWFS measurement and the previous DM commands from lag frames behind. Only the modes controlled by the leaky integrator are reconstructed in this step. At any point, up to 120000 frames (2 minutes at 1 kHz) of POL DM commands are available in the shared memory. 
 
\subsubsection{Predictive filter}

There are a few different flavors of predictive control implemented on the PyWFS RTC that all aim to minimize the same cost function: the linear minimum mean square error (LMMSE). We write the cost function in Eq.~\ref{eq:cost_function} where a unique predictive filter, $F^i$ is determined for each mode or in our case DM actuator. Our regressors are contained in the history vector $h(t)$ which are both spatial and temporal formed from our POL DM commands. Finally, the POL DM commands we want to predict are represented by $w_i$ at $\delta t$ steps into the future. 

\begin{equation}
    min_{F^i} < || F^ih(t) -w_i(t+\delta t) || ^2 >
    \label{eq:cost_function}
\end{equation}

In 2019, EOF was implemented on the PyWFS RTC in python such that the predictive filter was calculated in a batch sense by storing up to $l$ wavefront sensor measurements~\cite{Jensen_2019}. In this implementation, the cost function is solved in a regularized least-squares sense as shown in Eq.~\ref{eq:EOF} where $\alpha$ is the regularization parameter and $I$ is the identity matrix. $D$ and $P$ contain $l$ pairs of the true wavefront ($w_i$) and history vectors, $h$, respectively. 

\begin{equation}
    F^i= PD^T(DD^T+\alpha I)^{-1}
    \label{eq:EOF}
\end{equation}

The matrix multiplication of the predictive filter has now been implemented in the gpu on the PyWFS RTC, allowing us to predict at the speed of the system, an upgrade since 2019~\cite{Jensen_2019}. The EOF filter, using Eq.~\ref{eq:EOF} is first calculated offline in python and then passed to the shared memory once prediction is enabled. At the moment, the filter takes 9 seconds to calculate using 1 minute of telemetry ($l$=60000 frames). We have also updated the filter such that the number of temporal orders (how many previous measurements used to form $h$), $n$, can vary. Previously, we had been limited to a temporal order of 10 but now have the option to vary this parameter, however, doing so greatly increase the computation time at the moment to minute timescales.  

More recently, the recursive LMMSE (RLMMSE) solution\cite{vanKooten_2019} has also been implemented on the Keck RTC within python. The RLMMSE allows for the system to continuously update the predictive filter, effectively tracking any changes in the input distrubance (atmospheric turbulence). The solution to the cost function, Eq.~\ref{eq:cost_function}, can be written in terms of the auto-covariance of $h$ and the cross-covariance between the $w_i$ and $h$ ($C_{hh}$ and $c_{hw_i}$, respectively) as shown in Eq.~\ref{eq:batch_LMMSE}. Using this equation, we can calculate the batch solution by once again storing up data and estimating the auto-covariance and the cross-covariance matricies. We can also build-up our covariances using Eqs.~\ref{eq:auto_update} and~\ref{eq:cross_update}. In practice, the filter first runs in the background, once it has initially converged using a few thousand frames (e.g., $l$=O(1000)), the filter is then sent to the shared memory and prediction is enabled. The filter can then update recursively as fast as possible within the python implementation (each calculation takes roughly 0.02 seconds giving us an update rate of 50 Hz). 

\begin{equation}
F^i=C_{hh}^+c_{hw_i}    
\label{eq:batch_LMMSE}
\end{equation}

\begin{equation}
    \label{eq:auto_update}
    C_{hh}^+(t)=C_{hh}^+(t-1)-\frac{C_{hh}^+(t-1)h^T(t)C_{hh}^+(t-1)}{1+h^T(t)C_{hh}^+(t-1)h(t)}
\end{equation}

\begin{equation}
    \label{eq:cross_update}
    c_{hw_i}(t)=c_{hw_i}(t-1)+w_i(t)h^T(t)
\end{equation}

\subsubsection{Control Law}
Our current implementation of the predictive filter is a semi-open loop controller. The predicted wavefronts return the DM commands for correcting the full atmospheric turbulence, $V^{pred}(t)$. The predicted commands can be mixed with the integrator to generate the signal sent to the DM, $V(t)$. This configuration is shown in Eq.~\ref{eq:open_loop}. The integral controller (the second term) is the same as Eq.~\ref{eq:leaky_int}. Now the slopes encode the residuals of the predictor.  We tune the mixing parameter $m$ and the gain $g$ to have stable correction as well as maintain the option to have a pure predictor when $g$ is zero or $m$ is unity. 
\begin{equation}
    V(t)=mV^{pred}(t)+(1-m)((k)V^{int}({t-1})-g(CM \times S(t)-S_{ref}))
    \label{eq:open_loop}
\end{equation}


\section{Daytime testing}
\label{sec:daytime}
As outlined in Sec.~\ref{sec:controller} changes were made to the current predictive filer implementation and new features such as RLMMSE were added. In this section, we present results from our daytime tests on the Keck II AO bench including results from these new features that have been tested during the day. 

\begin{figure}[ht]
    \centering
    \includegraphics[width=\textwidth,trim={0 0cm 0cm 0cm},clip]{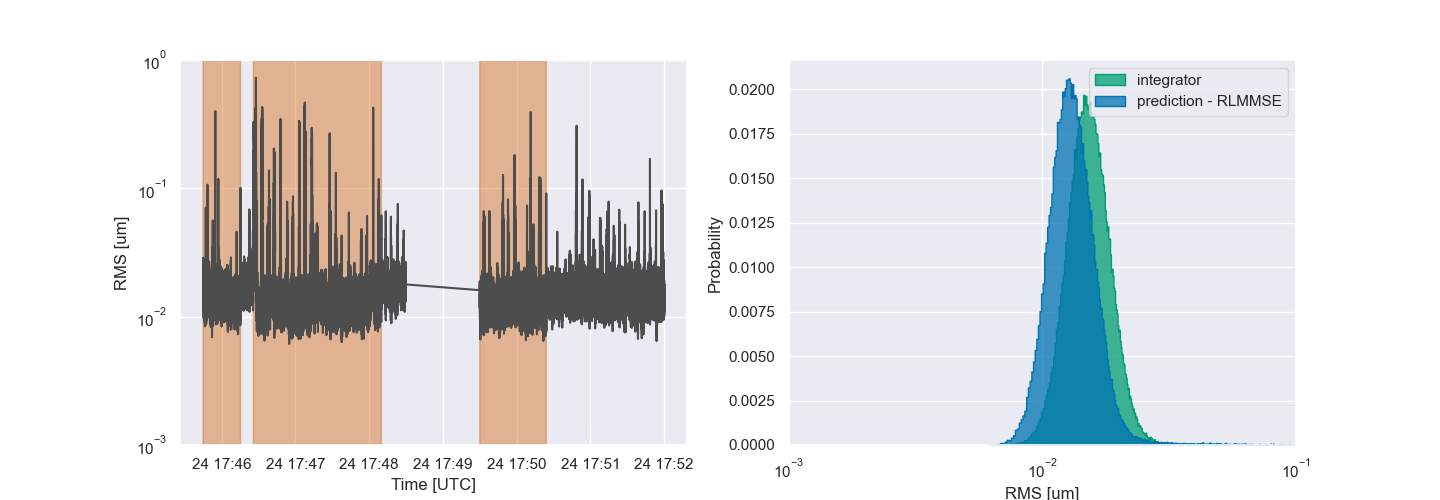}
    \caption{The left panel shows a time series of RMS wavefront error taken during the day on May 24 2021 while using the RLMMSE algorithm. The right panel shows the probability density functions for the predictor and integrator. This was taken while playing single layer turbulence on the DM with $r_0 =16cm$ and a wind speed of 15$m/s$. In both panels, the shaded orange regions indicate when prediction was turned on. }
    \label{fig:timeseries_RLMMSE}
\end{figure}
In Fig.~\ref{fig:timeseries_RLMMSE}, we show the results from testing the RLMMSE algorithm with single layer turbulence imposed on the DM on May 24, 2021. The RMS plotted is the residual wavefront error as measured by the PyWFS. To determine these values, the slopes are multiplied by the closed-loop command matrix (or on-sky matrix). Therefore, the RMS wavefront error does not include residuals from modes that are not controlled by the AO system. These errors of course will still influence the image quality of NIRC2. The modest improvement of 1.15 times shown in Fig.~\ref{fig:timeseries_RLMMSE} confirms that the implementation and the algorithm work as expected. More day time testing is needed to find the optimal parameters for the RLMMSE (such as the initial training time) as well as to improve the rate at which the RLMMSE updates. Currently, updating the filter is done within python and takes around 0.02 seconds. 

\begin{figure}[ht]
    \centering
    \includegraphics[width=\textwidth,trim={0 0cm 0cm 0cm},clip]{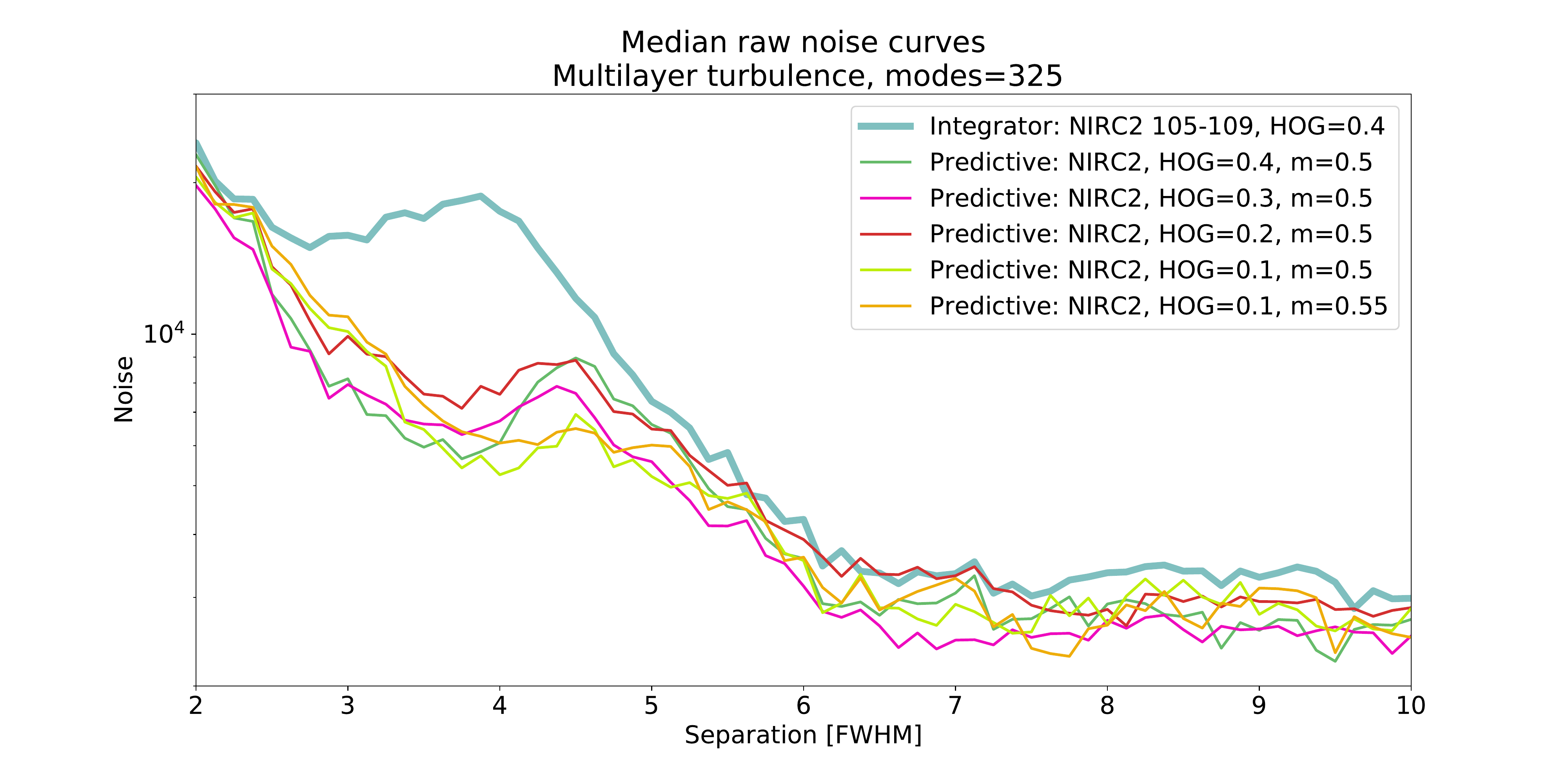}
    \caption{ Median noise curves for K-band coronagraphic NIRC2 images taken with 325 controllable modes. The integrator gain was set to 0.4 during the integrator dataset. During the predictive dataset, the integrator gain was reduced to 0.2. The mixing factor ($m$) varied slightly while we changed the HO gain. }
    \label{fig:daytime_contrast_modes325}
\end{figure}

\begin{figure}[ht]
    \centering
    \includegraphics[width=\textwidth,trim={0 0cm 0cm 0cm},clip]{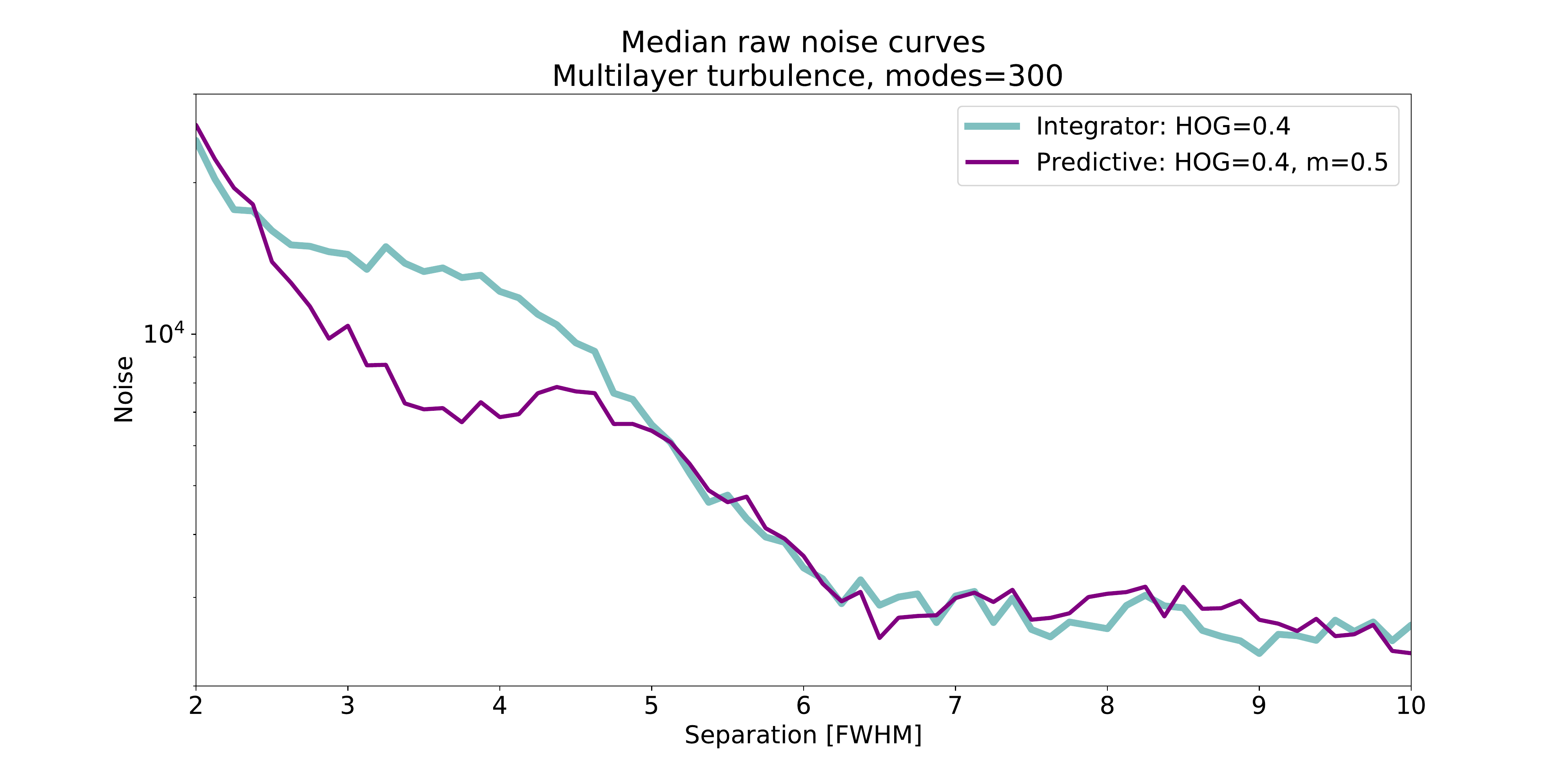}
    \caption{Median noise curves for K-band coronagraphic NIRC2 images taken with 300 controllable modes. The integrator gain was set to 0.4 during the integrator dataset and then reduced to 0.2 for the predictor. }
    \label{fig:daytime_contrast_modes300}
\end{figure}

During the day of July 27, 2021, we tested EOF under a variety of different configurations and focused on varying the number of modes and the HO gain while taking K-band coronagraphic images with NIRC2. We applied mutli-layer turbulence with the parameters outlined in the Keck Adaptive Optics Note No 303 (KAON303)~\cite{KAON303}. Setting $\alpha=1$, and training on 2 minutes of data, the EOF achieved good correction and provided an improvement over the nominal integrator configuration.  We plot the median raw noise curves for 325 controlled modes in Fig.~\ref{fig:daytime_contrast_modes325} and for 300 controlled modes in Fig.~\ref{fig:daytime_contrast_modes300} of the NIRC2 images throughout the daytime tests. These images show the contrast out to the DM control radius of 10$\lambda/D$. The noise as a function of separation was computed using the Vortex Image Processing (VIP~\cite{2017AJ....154....7G}) python package: a ring of FWHM-diameter circular apertures was constructed at each separation, and the noise was taken to be the standard deviation of the aperture sums at a given separation.  Figs.~\ref{fig:daytime_contrast_modes325} and \ref{fig:daytime_contrast_modes300} show that the predictor is able to perform better (by a factor of 3.5 for 300 modes and 4-5 for 325 modes) than the integrator at small separations (3-4 $\lambda/D$; where we expect predictive control to provide an improvement) under a variety of different system configurations. Figure~\ref{fig:daytime_contrast_modes325} shows that varying the HO gain for the predictor does not have a large impact on its performance. We, however, acknowledge that this behavior might be very different on-sky because here we are introducing turbulence using the DM itself, and therefore we do not have the same amount of HO spatial frequencies to correct as we would on-sky. Finally, comparing the two figures to each other we see that the integrator's performance appears to be more affected by the number of corrected modes than the predictor and that fewer modes here provide a better noise curve. 

\section{On-sky testing}
\label{sec:nighttime}

In this section, we present the results from two different engineering nights that allowed for on-sky testing of predictive control: the nights of June 20, 2021 and July 27, 2021 Hawaii Standard Time (HST).

\subsection{Night of June 20 2021}
On June 20, 2021, we received a quarter night at Keck for on-sky testing of predictive control. After some initial issues with the secondary mirror of the telescope, we were able to close the AO loop on TIC 125676063 -- a bright star with an H-band magnitude of 4.24. Due to the low SR being provided by the leaky integrator, we took non-coronagraphic images. We attribute the poor performance of the AO system during our observing to secondary mirror issues and the last minute requirement to revert to an older command matrix. 

\begin{figure}[ht]
    \centering
    \includegraphics[width=0.9\textwidth,trim={0 1.5cm 1cm 0cm},clip]{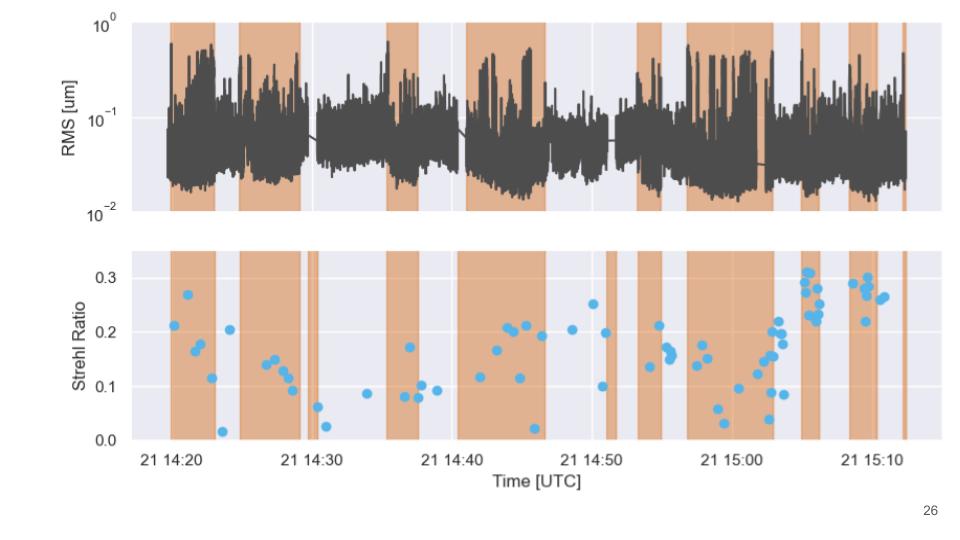}
    \caption{The top panel shows a time series RMS wavefront error for the entire testing period taken starting the night of June 20, 2021. The bottom panel shows the Strehl ratio as measured from the NIRC2 images. The shaded orange region indicates when prediction was turned on.  }
    \label{fig:timeseries_June20}
\end{figure}

\begin{figure}%
    \centering
    \subfloat[\centering Histogram of RMS wavefront error for frame lag 1.]{{\includegraphics[width=10.5cm,trim={0.5cm 9.35cm 15.5cm 0.4cm},clip]{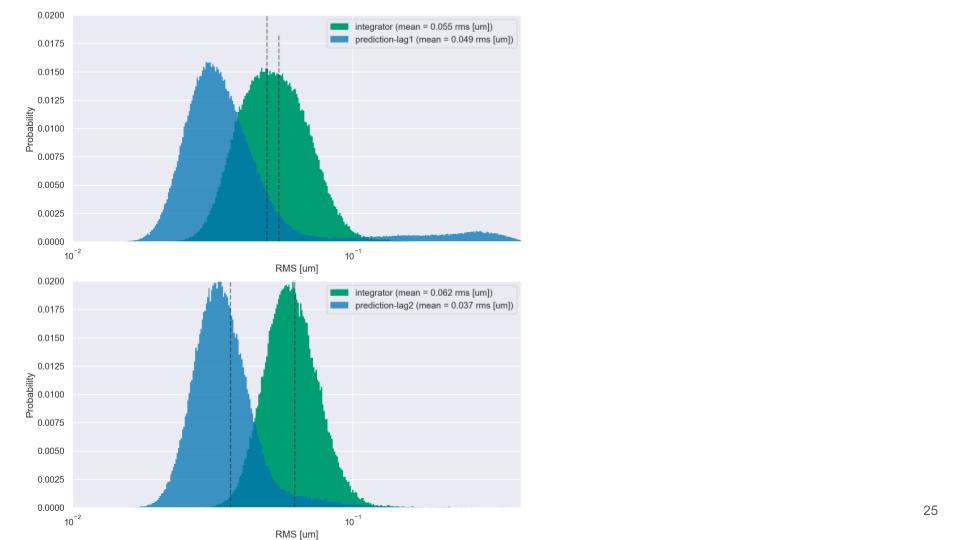}}}%
    \qquad
    \subfloat[\centering Histogram of RMS wavefront error for frame lag 2.  ]{{\includegraphics[width=10.5cm,trim={0.5cm 0cm 15.5cm 9.75cm},clip]{Figures/SPIE_2021_June20_histogram.jpg}}}%
   \caption{Analysis of night time tests on June 20 2021. The integrator data was taken before and after prediction was turned on and off. }%

    \label{fig:hist_June20}  %
\end{figure}

In Fig.~\ref{fig:timeseries_June20}, we show the RMS wavefront error for the entire night. We also show the SR as measured from NIRC2 images. Studying the time series of the RMS, we can clearly see the improvements provided by prediction. The time series also reveals the effects of all the changes made to the AO controller while observing.  The continued improvement of prediction in a shaded region is a result of the mixing factor and integrator gain being adjusted to find the optimal configuration. During this observation period, we also tested the effects of changing the assumed frame lag in the POL step. Figure~\ref{fig:hist_June20} shows the results from a lag of 1 and lag of 2. With a lag of 1 we are able to improve the performance, but the probability density function (PDF) is skewed with a tail tending towards larger RMS values. This tail is removed when a lag of 2 is used by the predictor. We also confirm the lag of 1.7 frames as determined by S. Cetre et al. 2018~\cite{Cetre_2018}. Further more, the lag of 2 PDF shows an improvement of 1.7 times in RMS wavefront error. While we do find that a lag of 2 is better for the predictor it is difficult to draw concrete conclusions when comparing to the integrator since its performance was impacted by secondary mirror issues. 


\subsection{Night of July 27 2021}
\begin{figure}
    \centering
    \includegraphics[width=\textwidth,trim={0cm 0cm 0cm 0cm},clip]{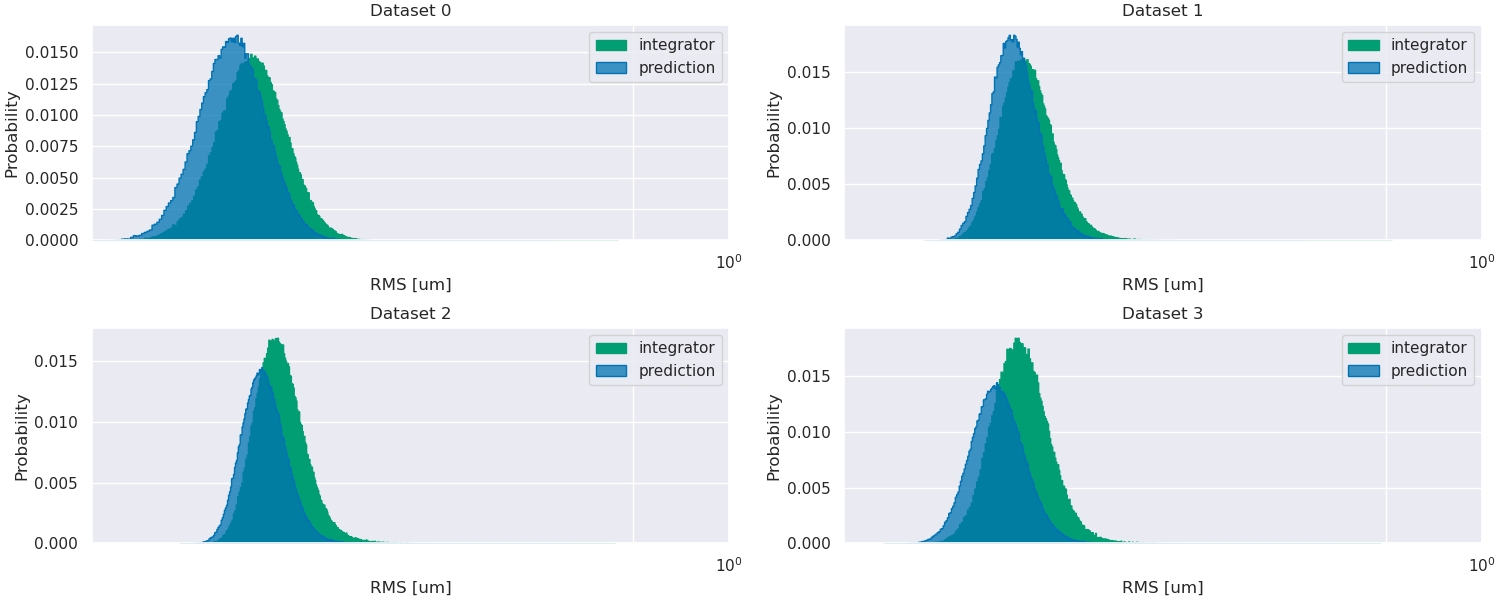}
    \caption{Probability density functions (PDF) for each of the different data sets separated by controller type. We compare the predictor with the integrator data taken just before the predictor is turned on. The different parameters for the data sets are provided in Tab.~\ref{tab:July28_parameters}.  }
    \label{fig:hist_July27}
\end{figure}

On the night of July 27, 2021, we received a half night of engineering time for predictive wavefront control tests. In this section, we present the results for one of our targets, HIP 117578. With an H-band magnitude of 6.1, HIP 117578 is bright enough to take advantage of the 1kHz frame rate of the PyWFS camera. We took NIRC2 images with the L-band coronagraph in place and make use of QACITS to keep the point-spread-function (psf) centered on the vortex coronagraph~\cite{QACITS}. 
\begin{figure}
    \centering
    \includegraphics[width=\textwidth,trim={1.5cm 18.14cm 0.5cm 3.1cm},clip]{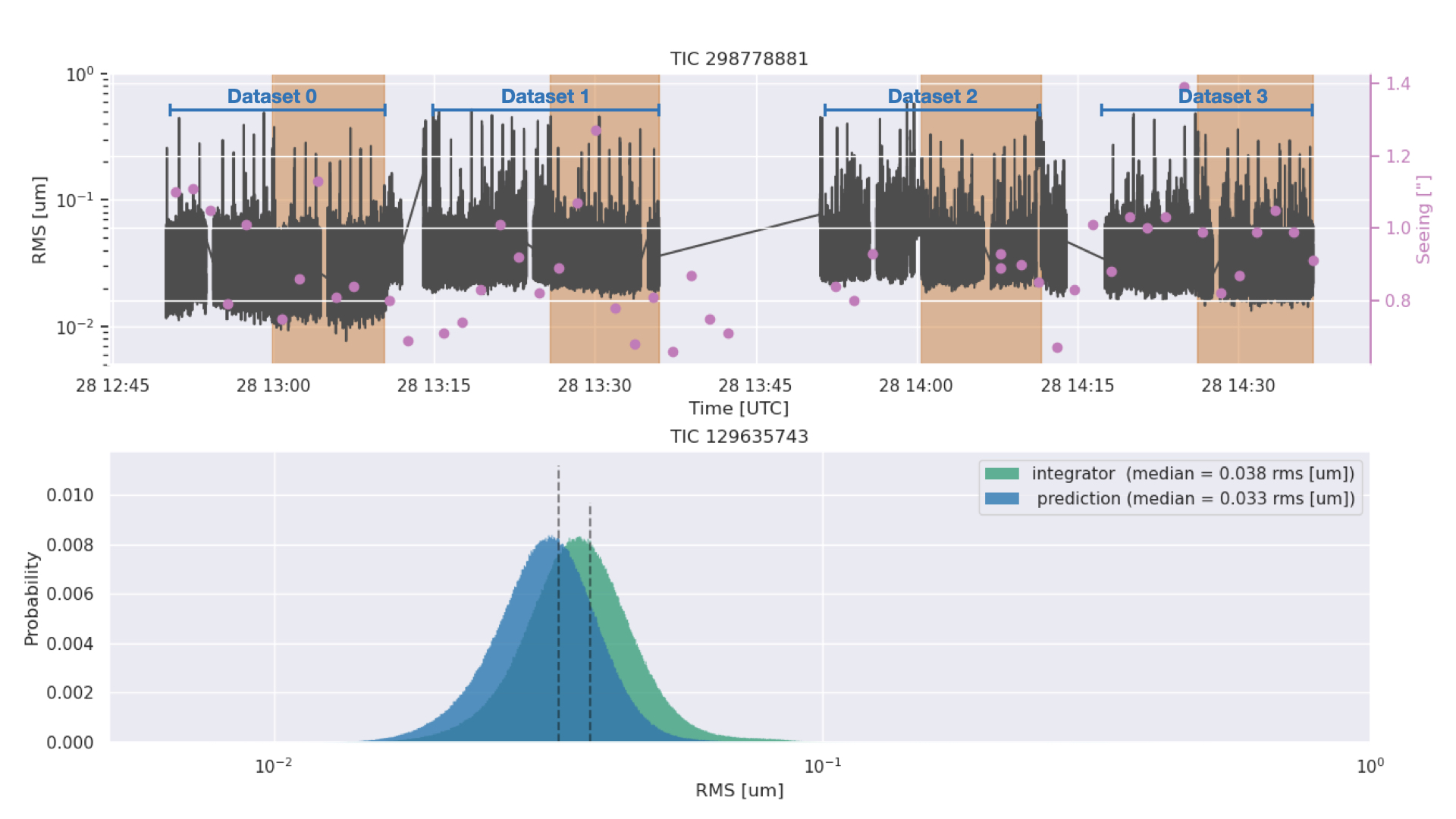}
    \caption{Time series of the RMS wavefront error for observations of HIP 117578 with the shaded regions indicating when prediction was turned on. The seeing is the DIMM as measured by CFHT (data source). We indicate four different datasets that were taken with various configurations as outlined in Tab.~\ref{tab:July28_parameters}.}
    \label{fig:timeseries_July27}
\end{figure}
Figure~\ref{fig:timeseries_July27} provides an overview of the observations with the RMS wavefront error being plotted as a function of time. We also plot the seeing as measured by the DIMM at the Canada France Hawaii Telescope (CFHT)~\cite{CFHT_weather}. The figure also indicates four data sets that were taken for this target. For each of the data set we configured the AO system as desired (see Tab.~\ref{tab:July28_parameters}) and then started the QACITS sequence which first runs an optimization sequence to center the vortex and then takes a pre-determined number of science images. Half way through this set of science images we switch from the integrator controller to the predictive controller. These full QACITS sequences, which include NIRC2 frames taken using the integrator followed by predictive control, constitute a ``dataset" (Tab. \ref{tab:July28_parameters}). The NIRC2 images were all taken with an exposure time of 0.3 seconds and 90 coadds at the full frame of 1024-by-1024 pixels. When we swap controllers, one to two NIRC2 frames are contaminated and hence dropped from this analysis. Therefore, we compare the predictor to integrator data that was taken just before the predictor data set. In Fig.~\ref{fig:hist_July27}, we plot the PDF of the RMS wavefront error for the four different data sets. These PDFs only include data that were taken during the NIRC2 exposures that are used for our contrast analysis. The ratio of the medians and standard deviations of the PDFs for each dataset are presented in Tab~\ref{tab:July28_parameters}. From these values, we see that the predictor provides an improvement in standard deviation (between a factor of 1.23 and 2.09 improvement in standard deviation) which suggests that the predictor provides a more stable AO correction than the integrator. Furthermore, for all but data set 0 the variation in seeing during the integrator and predictor exposures within a data set where similar. Meteorological data from CFHT~\cite{CFHT_weather} show that there was no change in the wind direction and little change in the wind speed during our observations. Therefore, it is unlikely that variation in seeing causes the improvement in standard deviation for the predictor. The improvement in AO stability is in agreement with results found by van Kooten et al. 2020~\cite{vanKooten_2020}.

\begin{table}[ht]
\centering
\caption{The ratios of the median and standard deviation RMS wavefront error for each data set along with the ratio of contrast (contrast gain) at a separation of 3 $\lambda/D$. The controller settings for each of the data sets are shown and the number of modes controlled for all data sets was 300 modes with the LO gain being set to 1. For the NIR2 images, the exposure was set to 0.3 seconds and the co-adds to 90 for all images. }
\label{tab:July28_parameters}
\small
\begin{tabular}{||c c c c c c c c c||} 
 \hline
 Dataset & $\frac{{Median_{int}}}{Median_{pred}}$ &$\frac{Std_{int}}{Std_{pred}}$&$\frac{Contrast_{int}}{Contrast_{pred}}$ & Integrator gain & Mixing factor ($m$)& HO gain & CoF & Order \\ [0.5ex] 
 \hline\hline
 0 & 1.17 & 1.45 & 1.51 & 0.15 &0.56 & 0 & 3 & 10 \\ 
 \hline
 1 & 1.11 & 2.09 & 1.57 & 0.3 & 0.5 & 0.35 & 3 & 10  \\
 \hline
 2 & 1.14 & 1.23 & 2.03 & 0.3 & 0.5 & 0 & 5 & 10  \\
 \hline
 3 & 1.19 & 1.31 & 0.22 & 0.3 & 0.57 & 0 & 5 & 20 \\
 \hline \hline
\end{tabular}

\end{table}

\begin{figure}%
    \centering
    \subfloat[]{{\includegraphics[width=8cm,trim={1cm 1cm 3cm 0.5cm},clip, page=1]{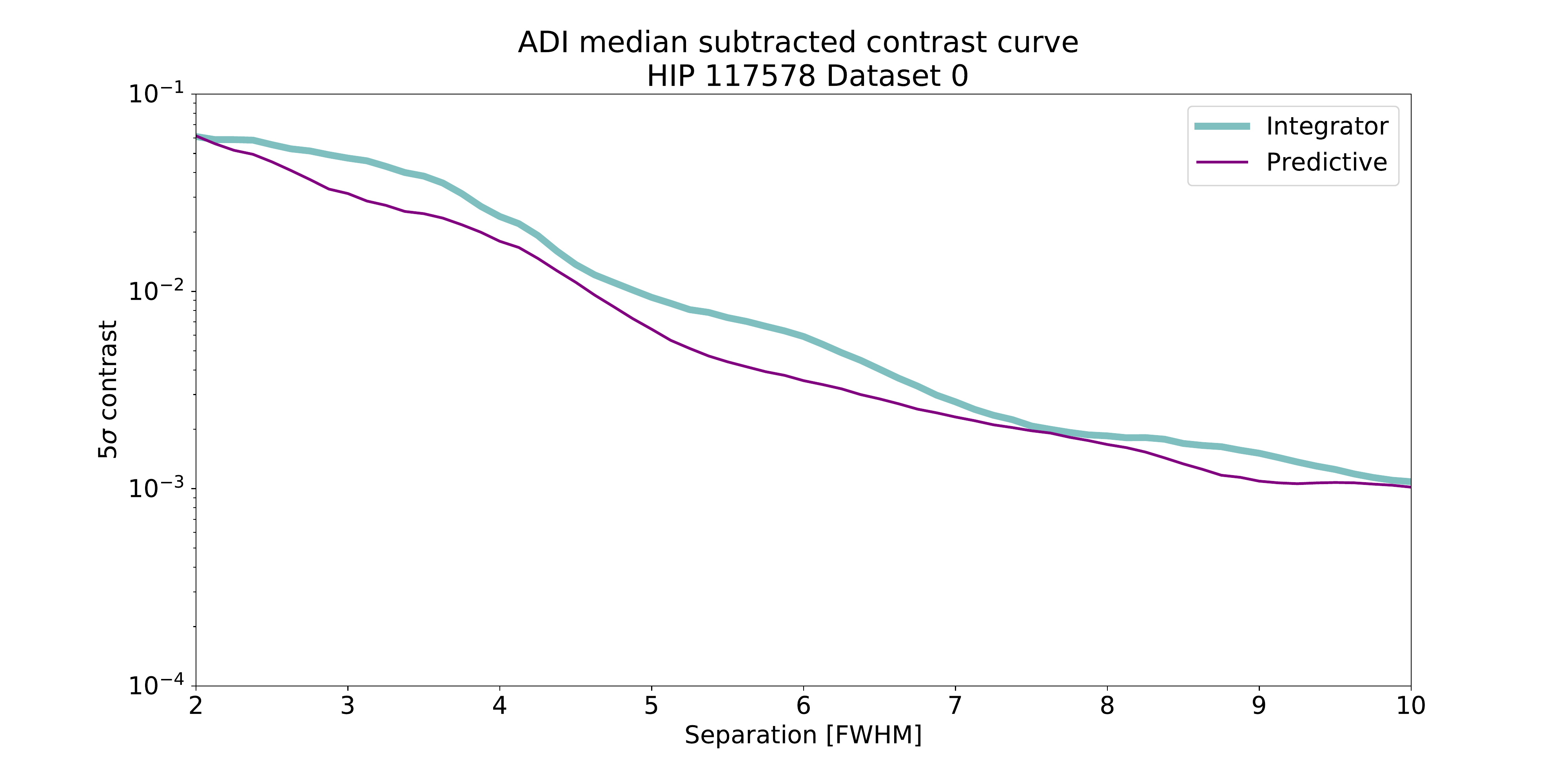}}}%
    \qquad
    \subfloat[ ]{{\includegraphics[width=8cm,trim={1cm 1cm 3cm 0.5cm},clip, page=2]{Figures/contrast_curves_ADI_TIC129635743.pdf}}}%
   
       \centering
    \subfloat[]{{\includegraphics[width=8cm,trim={1cm 1cm 3cm 0.5cm},clip, page=3]{Figures/contrast_curves_ADI_TIC129635743.pdf}}}%
    \qquad
    \subfloat[ ]{{\includegraphics[width=8cm, trim={1cm 1cm 3cm 0.5cm},clip,page=4]{Figures/contrast_curves_ADI_TIC129635743.pdf}}}%
   \caption{Angular differential imaging contrast curves for HIP 117578 taken on the night of July 27 2021 for the four different datasets as outline in Tab.~\ref{tab:July28_parameters}.  }%

    \label{fig:contrast_july28}  %
\end{figure}

We present the contrast curves for the four different data sets taken while observing HIP 117578 in Fig.~\ref{fig:contrast_july28}. The NIRC2 data was first bad pixel, flat-field, and sky corrected, as well as registered, using the automated pipeline described in Xuan et al.~2018\cite{2018AJ....156..156X}. We used VIP to de-rotate, median subtract, and median combine the data, modifying the number of images in each dataset such that the integrator and predictive components had the same total parallactic rotation. We then used VIP to create student-t corrected, algorithmic throughput corrected contrast curves, as detailed in Gomez Gonzalez et al.~2017\cite{2017AJ....154....7G}.  From the contrast curves we find a contrast gain of 2 for dataset 2 at a separation of 3~$\lambda/D$ as reported in Tab.~\ref{tab:July28_parameters}. In all of the datasets we see an improvement in contrast located at different separations. Figure~\ref{fig:contrast_gains_july28} in App.~\ref{app:gains} shows the contrast gain (ratio of the integrator and predictor contrast curves) for the four data sets, clearly showing the achieved improvement with predictive control. In data set 2 we see contrast gains close to 3 at separations of 4-6 $\lambda/D$ while we see more modest gains around 1.5 for data sets 0 and 3. From these contrast curves, we have repeatable improvement in contrast with EOF predictive control take over a period of 1.5 hours.

\section{Future work}
\label{sec:future}
Given the contrast improvement illustrated in Fig. \ref{fig:contrast_july28}, we have a clear path forward to improve the performance of the predictive control including: \begin{enumerate*}
\item implement predictive control in a true closed-loop configuration, 
\item determine the optimal parameters for both EOF and RLMMSE, 
\item match predictive control results to end-to-end simulations expanding on our integrator simulations that match on-sky contrast~\cite{Jensen-Clem_2021_PSI}, and
\item continue on-sky verification of predictive control under different conditions and guide star magnitudes. 
\end{enumerate*} 
Specifically, we outline below the next steps to operate predictive control in a true closed-loop configuration. Implementing integral action on the predicted atmospheric turbulence  will provide a more stable performance while also limiting the number of tunable parameters, further simplifying the use of the algorithm. 

In a closed-loop configuration the error measured by the PyWFS, $e(t)$, is the difference between the input disturbance $d(t)$ and the corrected phase given by the control signal at the last time step, $V(t-1)$, as shown in Eq.~\ref{eq:error}.

\begin{equation}
e(t)=d(t)-V(t-1)
\label{eq:error}
\end{equation}.

Assuming the PyWFS measures the true error signal perfectly we can write $e(t)$ as a function of the slopes as in Eq.~\ref{eq:error_slopes}. 

\begin{equation}
e(t)=CM \times S(t)
\label{eq:error_slopes}
\end{equation}
The predictive filter is predicting the full input disturbance, $d(t)$, that will occur when the commands are sent. We can plug Eq.~\ref{eq:error} into Eq.~\ref{eq:leaky_int} to get the closed-loop configuration of predictive control with integral action, Eq.~\ref{eq:pred_closed}. 
\begin{equation}
    V(t)=(k+g) V^(t-1)-gV(t)_{pred} +g (CM \times-S_{ref})
\label{eq:pred_closed}
\end{equation}

\section{Conclusions}
\label{sec:conclusions}

In this paper we present the latest results from predictive control on W. M. Keck II AO system. We show a repeatable improvement in contrast at small separations (3-5 $\lambda/D$) for both day time tests and on-sky tests. We find contrast gains of a factor of 1.5-3 on sky. Close inspection of the RMS wavefront error during the exposure of the NIRC2 images used for contrast analysis shows a reduction in the standard deviations of the distribution of wavefront errors for predictive control, revealing that prediction provides a more stable correction over these time periods. 

We also present: 
\begin{enumerate*}
\item a working implementation of RLMMSE during daytime tests,
\item a more optimized EOF predictor using a lag of 2 frames and $\alpha$ equal to one.
\item two different nights showing a reduction in RMS wavefront error with predictive control, and
\item a clear path to future controller upgrades to improve upon our predictive control implementation. 
\end{enumerate*}

\appendix    
\section{Contrast Gains}
\begin{figure}%
    \centering
    \subfloat[]{{\includegraphics[width=8cm,trim={1cm 1cm 3cm 0.5cm},clip, page=1]{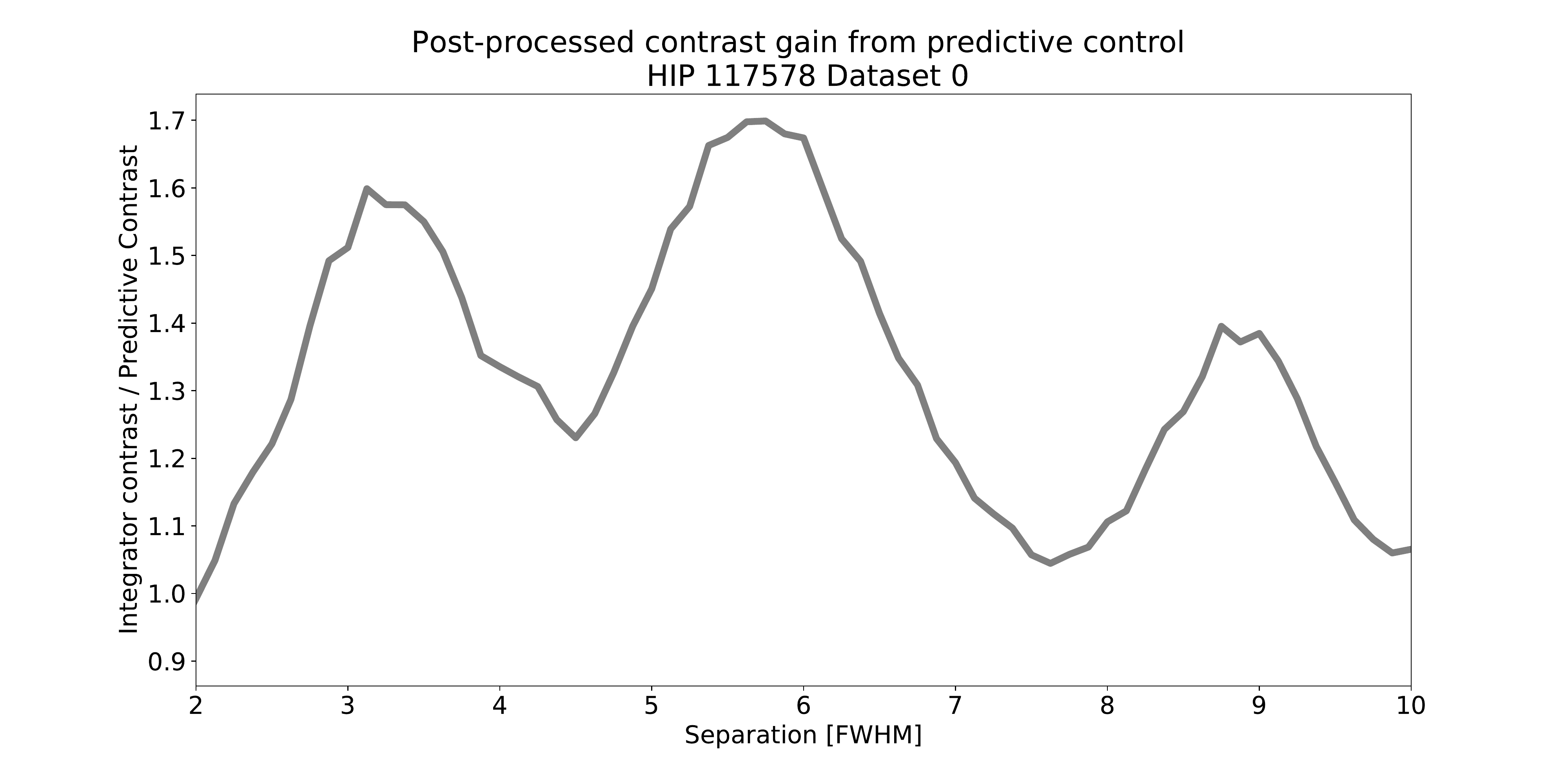}}}%
    \qquad
    \subfloat[ ]{{\includegraphics[width=8cm,trim={1cm 1cm 3cm 0.5cm},clip, page=2]{Figures/contrast_gain_TIC129635743.pdf}}}%
   
       \centering
    \subfloat[]{{\includegraphics[width=8cm,trim={1cm 1cm 3cm 0.5cm},clip, page=3]{Figures/contrast_gain_TIC129635743.pdf}}}%
    \qquad
    \subfloat[ ]{{\includegraphics[width=8cm, trim={1cm 1cm 3cm 0.5cm},clip,page=4]{Figures/contrast_gain_TIC129635743.pdf}}}%
   \caption{Contrast gains for HIP 117578 calculated by dividing the ADI contrast from the integrator by the ADI predictive contrast. Values greater than 1 demonstrate better performance with the predictive controller. The datasets correspond to the datasets in Tab.~\ref{tab:July28_parameters}}.%
    \label{fig:contrast_gains_july28}  %
\end{figure}

For further reference, we show the contrast gains achieved for the four different datasets in Fig.~\ref{fig:contrast_gains_july28}. These curves were calculated by dividing the contrast curves shown in Fig.~\ref{fig:contrast_july28}. 

\label{app:gains}

\acknowledgments 
The predictive wavefront control demonstration is funded by the Heising-Simons Foundation. The W. M. Keck Observatory is operated as a scientific partnership among the California Institute of Technology, the University of California, and the National Aeronautics and Space Administration. The Observatory was made possible by the generous financial support of the W. M. Keck Foundation. The near-infrared pyramid wavefront sensor (PyWFS) is supported by the National Science Foundation under Grant No. AST-1611623. The PyWSF camera was provided by Don Hall as part of his National Science Foundation funding under Grant No. AST 1106391.  This work benefited from the NASA Nexus for Exoplanet System Science (NExSS) research coordination network sponsored by the NASA Science Mission Directorate. The authors wish to recognize and acknowledge the very significant cultural role and reverence that the summit of Maunakea has always had within the indigenous Hawaiian community. We are most fortunate to have the opportunity to conduct observations from this mountain.

\bibliography{report} 

\begin{thebibliography}{10}

\bibitem{Bond_2020}
{Bond}, C.~Z., {Cetre}, S., {Lilley}, S., {Wizinowich}, P., {Mawet}, D.,
  {Chun}, M., {Wetherell}, E., {Jacobson}, S., {Lockhart}, C., {Warmbier}, E.,
  {Ragland}, S., {Alvarez}, C., {Guyon}, O., {Goebel}, S., {Delorme}, J.-R.,
  {Jovanovic}, N., {Hall}, D.~N., {Wallace}, J.~K., {Taheri}, M., {Plantet},
  C., and {Chambouleyron}, V., ``{Adaptive optics with an infrared pyramid
  wavefront sensor at Keck},'' {\em Journal of Astronomical Telescopes,
  Instruments, and Systems}~{\bf 6},  039003 (July 2020).

\bibitem{zurlo_2016}
{Zurlo, A.}, {Vigan, A.}, {Galicher, R.}, {Maire, A.-L.}, {Mesa, D.}, {Gratton,
  R.}, {Chauvin, G.}, {Kasper, M.}, {Moutou, C.}, {Bonnefoy, M.}, {Desidera,
  S.}, {Abe, L.}, {Apai, D.}, {Baruffolo, A.}, {Baudoz, P.}, {Baudrand, J.},
  {Beuzit, J.-L.}, {Blancard, P.}, {Boccaletti, A.}, {Cantalloube, F.}, {Carle,
  M.}, {Cascone, E.}, {Charton, J.}, {Claudi, R. U.}, {Costille, A.}, {de
  Caprio, V.}, {Dohlen, K.}, {Dominik, C.}, {Fantinel, D.}, {Feautrier, P.},
  {Feldt, M.}, {Fusco, T.}, {Gigan, P.}, {Girard, J. H.}, {Gisler, D.}, {Gluck,
  L.}, {Gry, C.}, {Henning, T.}, {Hugot, E.}, {Janson, M.}, {Jaquet, M.},
  {Lagrange, A.-M.}, {Langlois, M.}, {Llored, M.}, {Madec, F.}, {Magnard, Y.},
  {Martinez, P.}, {Maurel, D.}, {Mawet, D.}, {Meyer, M. R.}, {Milli, J.},
  {Moeller-Nilsson, O.}, {Mouillet, D.}, {Orign\'e, A.}, {Pavlov, A.}, {Petit,
  C.}, {Puget, P.}, {Quanz, S. P.}, {Rabou, P.}, {Ramos, J.}, {Rousset, G.},
  {Roux, A.}, {Salasnich, B.}, {Salter, G.}, {Sauvage, J.-F.}, {Schmid, H. M.},
  {Soenke, C.}, {Stadler, E.}, {Suarez, M.}, {Turatto, M.}, {Udry, S.},
  {Vakili, F.}, {Wahhaj, Z.}, {Wildi, F.}, and {Antichi, J.}, ``First light of
  the vlt planet finder sphere - iii. new spectrophotometry and astrometry of
  the hr9 exoplanetary system,'' {\em A\&A}~{\bf 587},  A57 (2016).

\bibitem{Cantalloube_2018}
{Cantalloube, F.}, {Por, E. H.}, {Dohlen, K.}, {Sauvage, J.-F.}, {Vigan, A.},
  {Kasper, M.}, {Bharmal, N.}, {Henning, T.}, {Brandner, W.}, {Milli, J.},
  {Correia, C.}, and {Fusco, T.}, ``Origin of the asymmetry of the wind driven
  halo observed in high-contrast images,'' {\em Astronomy \& Astrophysics}~{\bf
  620},  L10 (2018).

\bibitem{Guyon_2017}
{Guyon}, O. and {Males}, J., ``{Adaptive Optics Predictive Control with
  Empirical Orthogonal Functions (EOFs)},'' {\em ArXiv e-prints}~{\bf ArXiv}
  (July 2017).

\bibitem{Nem_2015}
{Jovanovic}, N., {Martinache}, F., {Guyon}, O., C.{Clergeon}, {Singh}, G.,
  {Kudo}, T., {Garrel}, V., {Newman}, K., {Doughty}, D., {Lozi}, J., {Males},
  J., {Minowa}, Y., {Hayano}, Y., {Takato}, N., {Morino}, J., {Kuhn}, J.,
  {Serabyn}, E., {Norris}, B., {Tuthill}, P., {Schworer}, G., {Stewart}, P.,
  {Close}, L., {Huby}, E., {Perrin}and S.~{Lacour}, G., {Gauchet}, L.,
  {Vievard}, S., {Murakami}, N., {Oshiyama}, F., {Baba}, N., {Matsuo}and
  J.~{Nishikawa}and M.~{Tamura}, T., {Lai}, O., {Marchis}, F., {Duchene}, G.,
  {Kotani}, T., and {Woillez}, J., ``The subaru coronagraphic extreme adaptive
  optics system: Enabling high-contrast imaging on solar-system scales,'' {\em
  Publications of the Astronomical Society of the Pacific}~{\bf 127},  890
  (Sept. 2015).

\bibitem{cacao}
Guyon, O., ``The compute and control for adaptive optics (cacao) real-time
  control software package,''
  https://indico.obspm.fr/event/57/contributions/208/attachments/171/189/cacao.pdf
  (2021).

\bibitem{Jensen_2019}
{Jensen-Clem}, R., {Bond}, C.~Z., {Cetre}, S., {McEwen}, E., {Wizinowich}, P.,
  {Ragland}, S., {Mawet}, D., and {Graham}, J., ``{Demonstrating predictive
  wavefront control with the Keck II near-infrared pyramid wavefront sensor},''
  ~{\bf 11117},  111170W (Sept. 2019).

\bibitem{Maaike_2017}
{van Kooten}, M., {Doelman}, N., and {Kenworthy}, M., ``Performance of ao
  predictive control in the presence of non-stationary turbulence,'' {\em
  AO4ELT5}~{\bf 5} (Sept. 2017).

\bibitem{vanKooten_2019}
van Kooten, M., Doelman, N., and Kenworthy, M., ``Impact of time-variant
  turbulence behavior on prediction for adaptive optics systems,'' {\em J. Opt.
  Soc. Am. A}~{\bf 36},  731--740 (May 2019).

\bibitem{vanKooten_2020}
{van Kooten, M. A. M.}, {Doelman, N.}, and {Kenworthy, M.}, ``Robustness of
  prediction for extreme adaptive optics systems under various observing
  conditions - an analysis using vlt/sphere adaptive optics data,'' {\em
  A\&A}~{\bf 636},  A81 (2020).

\bibitem{Haffert_2021}
{Haffert}, S.~Y., {Males}, J.~R., {Close}, L.~M., {Van Gorkom}, K., {Long},
  J.~D., {Hedglen}, A.~D., {Guyon}, O., {Schatz}, L., {Kautz}, M., {Lumbres},
  J., {Rodack}, A., {Knight}, J.~M., {Sun}, H., and {Fogarty}, K.,
  ``{Data-driven subspace predictive control of adaptive optics for
  high-contrast imaging},'' {\em arXiv e-prints}~{\bf ArXiv},  arXiv:2103.07566
  (Mar. 2021).

\bibitem{MagAOX}
{Males}, J.~R., {Close}, L.~M., {Miller}, K., {Schatz}, L., {Doelman}, D.,
  {Lumbres}, J., {Snik}, F., {Rodack}, A., {Knight}, J., {Van Gorkom}, K.,
  {Long}, J.~D., {Hedglen}, A., {Kautz}, M., {Jovanovic}, N., {Morzinski}, K.,
  {Guyon}, O., {Douglas}, E., {Follette}, K.~B., {Lozi}, J., {Bohlman}, C.,
  {Durney}, O., {Gasho}, V., {Hinz}, P., {Ireland}, M., {Jean}, M., {Keller},
  C., {Kenworthy}, M., {Mazin}, B., {Noenickx}, J., {Alfred}, D., {Perez}, K.,
  {Sanchez}, A., {Sauve}, C., {Weinberger}, A., and {Conrad}, A., ``{MagAO-X:
  project status and first laboratory results},'' ~{\bf 10703},  1070309 (July
  2018).

\bibitem{2018AJ....156..156X}
{Xuan}, W.~J., {Mawet}, D., {Ngo}, H., {Ruane}, G., {Bailey}, V.~P., {Choquet},
  {\'E}., {Absil}, O., {Alvarez}, C., {Bryan}, M., {Cook}, T., {Femen{\'\i}a
  Castell{\'a}}, B., {Gomez Gonzalez}, C., {Huby}, E., {Knutson}, H.~A.,
  {Matthews}, K., {Ragland}, S., {Serabyn}, E., and {Zawol}, Z.,
  ``{Characterizing the Performance of the NIRC2 Vortex Coronagraph at W. M.
  Keck Observatory},'' {\em The Astronomical Journal}~{\bf 156},  156 (Oct.
  2018).

\bibitem{L_band_vortex}
Castellá, B.~F., Serabyn, E., Mawet, D., Absil, O., Wizinowich, P., Matthews,
  K., Huby, E., Bottom, M., Campbell, R., Chan, D., Carlomagno, B., Cetre, S.,
  Defrère, D., Delacroix, C., Gonzalez, C.~G., Jolivet, A., Karlsson, M.,
  Lanclos, K., Lilley, S., Milner, S., Ngo, H., Reggiani, M., Simmons, J.,
  Tran, H., Catalan, E.~V., and Wertz, O., ``{Commissioning and first light
  results of an L'-band vortex coronagraph with the Keck II adaptive optics
  NIRC2 science instrument},'' ~{\bf 9909},  697 -- 710 (2016).

\bibitem{Bond_2018}
Bond, C.~Z., Wizinowich, P., Chun, M., Mawet, D., Lilley, S., Cetre, S.,
  Jovanovic, N., Delorme, J.-R., Wetherell, E., Jacobson, S., Lockhart, C.,
  Warmbier, E., Wallace, J.~K., Hall, D.~N., Goebel, S., Guyon, O., Plantet,
  C., Agapito, G., Giordano, C., Esposito, S., and Femenia-Castella, B.,
  ``{Adaptive optics with an infrared pyramid wavefront sensor},'' ~{\bf
  10703},  642 -- 652 (2018).

\bibitem{Cetre_2018}
{Cetre}, S., {Guyon}, O., {Bond}, C., {Chun}, M., {Mawet}, D., {Wizinowich},
  P., {Lockhart}, C., {Goebel}, S., and {Wetherell}, E., ``{A near-infrared
  pyramid wavefront sensor for Keck adaptive optics: real-time controller},''
  ~{\bf 10703},  1070339 (July 2018).

\bibitem{KAON303}
Neyman, C., ``Keck adaptive optics note 303: Atmospheric parameters for mauna
  kea,'' tech. rep., W. M. Keck Observatory (December 2004).

\bibitem{2017AJ....154....7G}
{Gomez Gonzalez}, C.~A., {Wertz}, O., {Absil}, O., {Christiaens}, V.,
  {Defr{\`e}re}, D., {Mawet}, D., {Milli}, J., {Absil}, P.-A., {Van
  Droogenbroeck}, M., {Cantalloube}, F., {Hinz}, P.~M., {Skemer}, A.~J.,
  {Karlsson}, M., and {Surdej}, J., ``Vip: Vortex image processing package for
  high-contrast direct imaging,'' {\em The Astronomical Journal}~{\bf 154},  7
  (July 2017).

\bibitem{QACITS}
Huby, E., Absil, O., Mawet, D., Baudoz, P., Castellà, B.~F., Bottom, M., Ngo,
  H., and Serabyn, E., ``{The QACITS pointing sensor: from theory to on-sky
  operation on Keck/NIRC2},'' ~{\bf 9909},  676 -- 685 (2016).

\bibitem{CFHT_weather}
CFHT, ``Mauna kea weather center,'' tech. rep.,
  http://mkwc.ifa.hawaii.edu/current/seeing/ (2021).

\bibitem{Jensen-Clem_2021_PSI}
Jensen-Clem, R., Hinz, P., van Kooten, M.~A., Chun, M., Fitzgerald, M.~P., Max,
  C., Mazin, B., Millar-Blanchaer, M., Guyon, O., Sallum, S., Skemer, A.,
  Stelter, R.~D., and Wang, J., ``{The Planetary Systems Imager Adaptive Optics
  System: a preliminary optical design and performance analysis of the PSI-Red
  AO system},'' ~{\bf 11823} (2021).

\end{thebibliography}
\bibliographystyle{spiebib} 

\end{document}